\documentclass[pra,twocolumn,10pt,aps,amsmath,amsfonts,superscriptaddress,floatfix]{revtex4-1}

\usepackage{graphicx}
\usepackage{dcolumn}
\usepackage{bm}
\usepackage{braket}
\usepackage{float}
\usepackage{amsmath}
\usepackage{hyperref}

\usepackage[usenames,dvipsnames]{color}

\newcommand{\W}{\mathbb{W}}

\renewcommand{\tt}{t}

\newcommand{\PP}{\mathbb{P}}
\newcommand{\LL}{\mathbb{L}}
\newcommand{\tW}{\tilde{\mathbb{W}}}
\newcommand{\tPP}{\tilde{\mathbb{P}}}

\newcommand{\N}{{\cal N}}

\makeatletter
\def\beq{\@ifstar{\@ifnextchar[{\@beqslabel}{\@beqsnolabel}}
{\@ifnextchar[{\@beqlabel}{\@beqnolabel}}}
\def\@beqlabel[#1]{\begin{equation}\label{#1}}
\def\@beqnolabel{\begin{equation}}
\def\@beqslabel[#1]{\begin{equation*}\label{#1}}
\def\@beqsnolabel{\begin{equation*}}
\def\eeq{\@ifstar{\end{equation*}}{\end{equation}}}
\makeatother

\newcommand{\punc}[1]{\;\text{#1}}

\newcommand{\reffiga}[2]{Fig.~\ref{#1}(#2)}

\newcommand{\vertdimer}{\textsf{II}}
\newcommand{\horidimer}{\rotatebox{90}{\textsf{II}}}

\newcommand{\ee}{e}

\newcommand{\sub}[1]{_{\text{#1}}}

\begin{document}

\title{A Deep Learning Functional Estimator of Optimal Dynamics for Sampling Large Deviations}

\author{Tom H.E. Oakes}
 \email{tom.oakes@nottingham.ac.uk}
\affiliation{School of Physics and Astronomy,  
University of Nottingham, Nottingham NG7 2RD, United Kingdom}
\affiliation{Centre for the Mathematics and Theoretical Physics of Quantum Non-equilibrium Systems, University of Nottingham, Nottingham NG7 2RD, UK}
\author{Adam Moss}
\affiliation{School of Physics and Astronomy,  
University of Nottingham, Nottingham NG7 2RD, United Kingdom}
\author{Juan P. Garrahan}
\affiliation{School of Physics and Astronomy,  
University of Nottingham, Nottingham NG7 2RD, United Kingdom}
\affiliation{Centre for the Mathematics and Theoretical Physics of Quantum Non-equilibrium Systems, University of Nottingham, Nottingham NG7 2RD, UK}

\date{\today}

\begin{abstract}
In stochastic systems, numerically sampling the relevant trajectories for the estimation of the large deviation statistics of time-extensive observables requires overcoming their exponential (in space and time) scarcity. The optimal way to access these rare events is by means of an auxiliary dynamics obtained from the original one through the so-called ``generalised Doob transformation''. While this optimal dynamics is guaranteed to exist its use is often impractical, as to define it requires the often impossible task of diagonalising a (tilted) dynamical generator. While approximate schemes have been devised to overcome this issue they are difficult to automate as they tend to require knowledge of the systems under study. 
Here we address this problem from the perspective of deep learning. We devise an iterative semi-supervised learning scheme which converges to the optimal or Doob dynamics with the clear advantage of requiring no prior knowledge of the system. We test our method in a paradigmatic statistical mechanics model with non-trivial dynamical fluctuations, the fully packed classical dimer model on the square lattice, showing that it compares favourably with more traditional approaches. We discuss broader implications of our results for the study of rare dynamical trajectories.
\end{abstract}

\maketitle

\section{Introduction}
\label{sec:intro}

In this paper we consider the problem of efficiently sampling rare trajectories of stochastic systems. A natural approach to studying stochastic dynamics is by means of trajectory ensemble methods, e.g.~\cite{Bolhuis2002,Merolle2005,Lecomte2007,Garrahan2007,Maes2008,Garrahan2009,Hedges2009,Baiesi2009,Jack2010,Pitard2011,Bodineau2012,Speck2012,Chetrite2013,Espigares2013,Budini2014,Chetrite2015,Jack2015,Bertini2015b} (for reviews see \cite{Eckmann1985,Ruelle2004,Gaspard2005,Touchette2009,Garrahan2018,Jack2019}). For the statics, the ensemble method of equilibrium statistical mechanics \cite{Chandler1987}
deals with sets of configurations or microstates and their probabilities of occurring under equilibrium conditions, and connects their statistical properties to macroscopically observable phenomena. The straightforward generalisation to dynamics 
\cite{Eckmann1985,Ruelle2004,Gaspard2005,Touchette2009,Garrahan2018,Jack2019}
is to replace microstates by stochastic trajectories, defining the ensemble in terms of all possible trajectories of the dynamics and the probabilities of them occurring under stochastic evolution. In the dynamical setting, order parameters (meaning macroscopic quantities that classify physical behaviour) correspond to observables of the whole trajectories, while the large size limit involves also the long time limit. Just like in the static case, in such ``thermodynamic'' regime large deviation (LD) principles may apply \cite{Eckmann1985,Ruelle2004,Gaspard2005,Touchette2009,Garrahan2018,Jack2019}, so that the statistics of observables become encoded in LD functions that play the role of free-energies for the dynamics (see below for definitions).

The usual setting is of a system with Markovian dynamics defined via its evolution operator 
(such as the exponential of a Markov generator for continuous-time dynamics), and one or more trajectory observables that one wishes to study \cite{Eckmann1985,Ruelle2004,Gaspard2005,Touchette2009,Garrahan2018,Jack2019}. As usual, physical intuition about the problem at hand will suggest that these are appropriate dynamical order parameters to describe the phenomenon of interest. For example they could correspond to macroscopic (particle or energy) currents for driven problems such as exclusion processes, or dynamical activities for systems with slow dynamics such as glasses. The aim is to calculate the probability distributions of these dynamical quantities. These distributions are ``physical'' in the sense that they are in principle observable with access to enough statistics of the dynamics. 

In practice, accessing the LD statistics of dynamical observables is challenging in general
\cite{Giardina2006,Cerou2007,Lecomte2007b,Gorissen2009,Giardina2011,Nemoto2016,Ray2018,Ray2018b,Klymko2018,Ferre2018,Banuls2019,Helms2019,Jacobson2019}. The extensive nature of the observables makes them exponential (in system size and time) concentrated around their typical or average value. That is, to observe trajectories other than typical is exponentially hard if sampled with the dynamics that generates them. Alternatively, the LD statistics is encoded in the spectrum of a deformation - or tiltings (see below for definitions) - of the evolution operator, but diagonalising such tilted operators is not possible except when the state space is of modest size. 

Many techniques have been developed to overcome these difficulties. Most of these approaches attempt to sample rare long-time trajectories efficiently by ameliorating their exponential scarcity within the original dynamics. Two popular methods are based either on population dynamics, such as cloning or splitting \cite{Giardina2006,Cerou2007,Lecomte2007b,Giardina2011}, or on importance sampling in trajectory space, such as as transition path sampling (TPS)
\cite{Bolhuis2002,Hedges2009}.
While these tame some of the exponential cost, further sampling improvements are needed to computational tractability in many systems of interest. Recently various schemes based on umbrella sampling in trajectory space have been shown to be useful
\cite{Ray2018,Ray2018b,Klymko2018}, together with related adaptive schemes based on optimal control 
\cite{Nemoto2016,Ferre2018,Das2019}
(more on both of these below). An alternative approach is via variational approximations, for example directly at the level of LD rate functions \cite{Jacobson2019}, or by using tensor networks to find leading eigenvectors of tilted generators \cite{Gorissen2009,Banuls2019,Helms2019}. 

While efficient for many problems of interest, the methods above often require some information on the problem being studied. Improving cloning or TPS with trajectory umbrella sampling implies the choice of an alternative sampling dynamics. While an optimal one is known to exist (via the so-called generalised Doob transformation \cite{Jack2010,Chetrite2015,Carollo2018}, see below), to find it requires solving the problem in the first place. Adaptive methods \cite{Nemoto2016,Ferre2018} attempt to find an approximation to this optimal dynamics, but for systems of interest a suitable parameterisation needs to be selected, and this if also problem dependent. Alternatives such as those based on variational tensor networks \cite{Gorissen2009,Banuls2019,Helms2019}, which in principle are generic, are in practice often restricted to one dimensional problems with short range interactions. 

Here we address the problem of finding in an automatic manner - i.e.\ without the usual physical input that goes into the parameterisation - an efficient auxiliary dynamics, close to the optimal Doob dynamics, for sampling rare trajectories in the LD regime. We address this problem from the perspective of deep learning \cite{Goodfellow2016,Mehta2019}, since this appears to be an unsupervised learning (USL) task of unknown features extractable from fluctuations of the original system dynamics. Many sampling techniques, like Monte Carlo (MC), rely on a form of encoding the configuration into eihter a unnique descriptor or some macroscopic value. The task of converting an cofiguration (or image) into a single value which encodes the likelyhood of that configuration from a given distribution is clearly a deep learning poblem \cite{kashiwa2019, tanaka2017, efthymiou2019}. What we know from studies like \cite{kashiwa2019} is that Neural Networks (NN) can encode more of these physical features than merely a label of which distribution the sample was taken from. Therefore given the right architecture and learning task a NN can be used to condense a configuration into a multitude of different parameters. In the past few years, there have been aproaches that take advantage of these NN's to speed up equlibration via sampling methods such as MC \cite{Huang2017a, Shen2018, Nagai2017, Huang2017b}. This type of approach to MC modification is perfectly framed for finding the auxiliary dynamics needed for the LD regime, using only a configuration as input and an appropriate learning process.

We focus for concreteness on a paradigmatic system with rich dynamics, namely, the fully packed classical dimer model (CDM) on the square lattice \cite{Fisher1961,Henley2010}, which is known to display a range of interesting dynamical LD behaviour \cite{Oakes2018}. We construct an iterative USL scheme which in progressively approximates with increased precision the optimal Doob dynamics for sampling. The clear advantage of our approach is that it does not require any a priori knowledge of the system. For the CDM we show that our method compares favourably with more standard approaches.

The paper is organised as follows. In Sect.~\ref{SecLD} we review the main concepts about trajectory ensembles and dynamical large deviations. In Sect.~\ref{Secmodel} we describe the classical dimer model which we use as a testbed for our ideas and describe the issues relating to sampling rare trajectories. Sections \ref{sec:NN} and \ref{sec:Unsuper} present our main results. In Sect.~\ref{sec:NN} we consider the problem of approximating an optimal dynamics for LD sampling via neural networks and supervised learning. It established the main principles and the appropriate architecture for the NNs. Subsequently, in Sect.~\ref{sec:Unsuper} we generalise the methods via semi-supervised learning to make them scalable and thus applicable to systems for which exact diagonalisation is not possible. Finally in Sect.~\ref{Secconc} we present our conclusions and outlook.

\section{Trajectory ensembles and dynamical large deviations}
\label{SecLD}

\subsection{Continuous-time Markov dynamics and trajectories}

For concreteness, we will focus on systems with stochastic Markovian dynamics in continuous time. The issues we discuss, however, are easily generalisable to discrete time dynamics. As a general setting we consider a system with configurations $\{ x_1, \cdots, x_\N \}$ where $\N$ indicates the size of configuration space (for example $\N = 2^N$ for a system of $N$ Ising spins). 

The master equation for the evolution of the probability over configurations can be written in general as
\beq[ME]
\partial_\tt \ket{P_\tt} = \W \ket{P_\tt} ,
\eeq
where $\left\{ \ket{x} \right\}$ is an orthonormal configuration basis, $\ket{P_\tt}$ is the probability vector, 
\beq[Pt]
\ket{P_\tt} = \sum_c P_\tt(x) \ket{x} ,
\eeq
and $P_\tt(x)$ the probability of configuration $x$ at time $\tt$. The Markov generator of the dynamics (sometimes also called master operator) is given by 
\beq[W]
\W = \sum_{x,x'\neq x} w_{x \rightarrow x'} \ket{x'} \bra{x} - \sum_{x} R_{x} \ket{x} \bra{x} ,
\eeq
The positive terms are off-diagonal and encode the possible transitions ${x \rightarrow x'}$ and their rates $w_{x \rightarrow x'}$. The negative terms are diagonal, with $R_{x}$ the escape rate from configuration $x$, $R_{x} = \sum_{x'\neq x} w_{x \rightarrow x'}$. The form \eqref{W} guarantees probability conservation: the largest eigenvalue of $\W$ is zero and its left eigenvector is the {\em uniform} (or ``flat'') state: 
\begin{align}
\label{pcons}
\bra{-} \W &= 0\punc,
&
\bra{-} &= \sum_x \bra{x}\punc.
\end{align}
In many cases, the dynamics generated by \eqref{W} also has a stationary state,
\begin{align}
\label{ss}
\W \ket{\rm ss} &= 0 \punc,
&
\ket{\rm ss} &= \sum_x P_{\rm ss}(x) \ket{x}\punc ,
\end{align}
where $P_{\rm ss}(x)$ denotes the (time independent) stationary state probability over configurations.

The generator \eqref{W} defines a continuous time Markov chain. Samples from such dynamics are stochastic trajectories, corresponding to a sequence of configurations and jumps between them at random times, 
\beq[traj]
\omega_\tt = (x_0 \to x_{\tt_1} \to \ldots \to x_{\tt_K})\punc,
\eeq
where $\tt_i$ ($i=1,\ldots,K$) indicate the times at which jumps between configurations occur, 
with time ordering $0 < \tt_1 < \cdots < \tt_K < \tt$, and we use the symbol $\omega_\tt$ to label a trajectory of total time extension $\tt$. Between jumps the configuration does not change unchanged, meaning that from the time of the last jump, $\tt_K$, and the final time $\tt$, the configuration in \eqref{traj} remains as $x_{\tt_K}$. 

At given conditions, the set of all possible trajectories $\omega_\tt$ generated by dynamics for, say, fixed total time extent $\tt$, together with their probability of occurring, $\pi(\omega_t)$, defines the (original or untilted, see below) ensemble of trajectories.

\subsection{Dynamical large deviations}

We are interested in the statistics of trajectory observables, i.e., functions of the whole trajectory which are extensive both in time and size of the system. One of the simplest is the {\em dynamical activity} 
\cite{Garrahan2007,Lecomte2007,Baiesi2009,Maes2019} defined as the number of configuration changes in a trajectory, $K(\omega_\tt)$. The distribution of these dynamical observables is obtained by contraction 
from that of the trajectories
\beq[PK]
p_\tt(K) = \sum_{\omega_\tt} \pi(\omega_\tt) \delta[K(\omega_\tt) - K] \punc.
\eeq
For long times $p_\tt(K)$ obeys a large deviation (LD) principle
\beq[PK2]
p_\tt(K) \sim e^{t \varphi(K/t)} \punc,
\eeq
where the (extensive in space) function $\varphi(k)$ is often called the rate function, and plays the role of an entropy density for trajectories. 
One can alternatively consider the statistics of $K$ in terms of its moment generating function (MGF) 
\cite{Eckmann1985,Ruelle2004,Gaspard2005,Touchette2009,Garrahan2018,Jack2019}, 
\beq[Zs]
Z_\tt(s) 
= \sum_K P_\tt(K) \ee^{-s K}
= \sum_{\omega_\tt} \pi(\omega_\tt) \ee^{- s K(\omega_\tt)} , 
\eeq
which also has a LD form at long times 
\beq[Zs2]
Z_\tt(s) \sim \ee^{\tt \theta(s)} .
\eeq
The scaled cumulant generating function(SCGF) $\theta(s)$ is a free-energy (per unit time, but extensive in system size) for trajectories, while $Z_\tt(s)$ is the corresponding trajectory partition sum. As in static thermodynamics, the rate function and SCGF are connected via a Legendre transform \cite{Eckmann1985,Ruelle2004,Gaspard2005,Touchette2009,Garrahan2018,Jack2019},
\beq[LT]
\theta(s) = - \min_{k} \left[ \varphi(k) + s k \right] .
\eeq

\subsection{Exponential tilting and biased ensembles}

The trajectory partition function \eqref{Zs} encodes an exponential tilting of the probabilities of trajectories. That is, it is the normalisation of a reweighing of the probability of a trajectory through the value of the observable $K$, 
\beq[sens]
\pi_s(\omega_\tt) = \frac{\pi(\omega_\tt) \ee^{- s K(\omega_\tt)}}{Z_\tt(s)} . 
\eeq
These tilted probabilities correspond to a biased trajectory ensemble, where averages of arbitraty trajectory observables are given by 
\begin{align}
\braket{A}_s = \frac{\braket{A \ee^{- s K}}}{Z_\tt(s)} 
&= \frac{\sum_{\omega} A(\omega) \, \pi(\omega) \ee^{- s K(\omega)}}{Z_\tt(s)}
\nonumber \\
&= \sum_{\omega} A(\omega) \, \pi_{s}(\omega)
 . 
\label{avs}
\end{align}
At $s=0$ these correspond to the averages under the process dynamics. For $s \neq 0$, they do not, but by tuning $s$ the exponential tilting allows to (formally) explore the properties of atypical trajectories. 

The dynamical partition sum \eqref{Zs} can be written in terms of a ``transfer matrix'',
\beq[Zss]
Z_\tt(s) = \bra{-} \ee^{\tt {\mathbb W}_s} \ket{\text{i}} \punc,
\eeq
where the probability vector $\ket{\text{i}}$ represents the distribution from which the initial state is drawn. The {\em tilted generator} ${{\mathbb W}_s}$ is a deformation of the original dynamical generator \cite{Eckmann1985,Ruelle2004,Gaspard2005,Touchette2009,Garrahan2018,Jack2019}. For the specific case of tilting against the activity (generalisations are straightforward) we have 
\beq[Ws]
\W_s = \ee^{-s} \sum_{x,x'\neq x} w_{x \rightarrow x'} \ket{x'} \bra{x} - \sum_{x} R_x \ket{x} \bra{x} \punc.
\eeq

The SCGF is the largest eigenvalue of $\W_{s}$, 
\begin{align}
\label{rls}
\W_{s} \ket{r_{s}} = \theta(s) \, \ket{r_{s}} \, , \;\;\;
 \bra{\ell_{s}} \W_{s} = \theta(s) \, \bra{\ell_{s}} \punc,
\end{align}
where $\ket{r_{s}}$ and $\bra{\ell_{s}}$ are the corresponding right and left eigenvectors.

\subsection{Basic considerations on sampling exponentially tilted ensembles}
\label{basic}

The spectral structure \eqref{rls} provides an important simplification. It converts the problem of estimating the LD statistics into one of minimising an operator - something that is routine in physics for example in the context of finding ground states of quantum Hamiltonians.
However, diagonalising $\W_{s}$ is often impractical in many-body systems once size exceeds modest values. In most cases of interest it is therefore necessary to resort to numerical approximation schemes. 

Consider the problem of calculating averages such as \eqref{avs}. Doing so using the original dynamics, \eqref{W}, is extremely impractical, as the exponential factor in the first line of \eqref{avs} makes convergence exponentially (in space and time) hard. Ideally one would like to sample $\pi_{s}(\omega)$ directly, but there is no easy way to generate trajectories compatible with \eqref{sens} starting from the known dynamics \eqref{W}. 

The first step in addressing the sampling problem is to account for the exponential factor in the first line of \eqref{avs} in a systematic manner, by means of importance sampling in trajectories. One such scheme is transition path sampling (TPS), see discussion below. It amounts to doing a biased random walk in trajectory space guaranteed to converge asymptotically to the biased trajectory ensemble \eqref{sens}. TPS will form the basis of the numerical scheme we develop below.

\begin{figure}
\begin{center}
\includegraphics[width=\columnwidth]{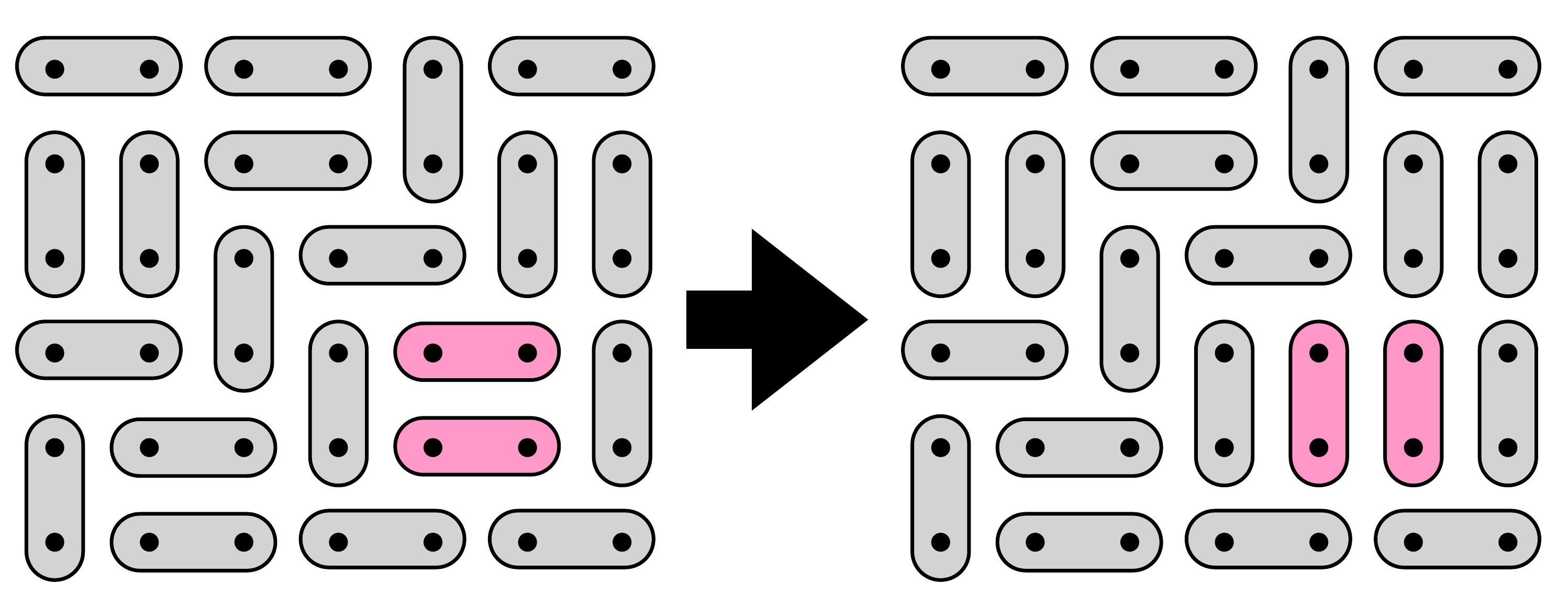}
\end{center}
\caption{An example of a fully packed classic dimer configuration showing how the local plaquette flip dynamics changes the configuration. Where the highlighted (pink) pair of dimers, or plaquette, undergo the plaquette flip dynamics which leaves all other dimers unchanged.}
\label{fig:cdm}
\end{figure}

\section{Classical dimer model and trajectory sampling}
\label{Secmodel}

\subsection{Model}
\label{SecCDM}

For concreteness we will consider the dynamics of a paradigmatic statistical mechanics model, that of fully packed dimers on a square lattice - with periodic boundary conditions - or classical dimer model (CDM)
\cite{Fisher1961,Henley2010,Oakes2018}. Figure 1 sketches the model. It comprises of dimers which occupy two adjacent sites on a square lattice (with periodic boundary conditions) under the conditions that all sites are covered and dimers do not overlap. For dynamics with local changes to the configuration $x$ which ``flip'' a pair of neighbouring parallel dimers (or {\em plaquette}), between horizontal to vertical and visa versa, without having any direct effect of the rest of the configuration, see fig.~1. 

The stochastic generator, cf.~\eqref{W}, for the CDM can be written schematically
\beq[Wcdm]
\W = \sum(\ket{\vertdimer} \bra{\horidimer} + \ket{\horidimer} \bra{\vertdimer}) - \sum(\ket{\vertdimer} \bra{\vertdimer} + \ket{\horidimer} \bra{\horidimer})\punc ,
\eeq
where the first sum contains all the non-diagonal elements that give rise to the plaquette transitions, while the second sum is diagonal and just counts them. Both the right eigenvector with eigenvalue zero (i.e.\ the equilibrium states) and the corresponding left eigenvector, Eqs.~\eqref{pcons}-\eqref{ss},  are given up to normalisation by the flat state over all allowed dimer configurations. As such, the dynamics generated by \eqref{Wcdm} is an ``infinite temperature'' dynamics, i.e., one where the stationary state is one of maximum entropy. 

The CDM is known \cite{Oakes2018} to have a LD first order transition when tilted by the dynamical activity, in this case given by the number of plaquette flips in a trajectory. This transition occurs at $s=0$ in the large size limit. The corresponding tilted generator, cf.~\eqref{Ws}, reads
\beq[Wcdms]
\W_s = e^{-s} \sum(\ket{\vertdimer} \bra{\horidimer} + \ket{\horidimer} \bra{\vertdimer}) - \sum(\ket{\vertdimer} \bra{\vertdimer} + \ket{\horidimer} \bra{\horidimer})\punc ,
\eeq
and the transition manifests in a first-order singularity in the SCGF $\theta(s)$ at $s=0$ in the large size limit.

\subsection{Transition path sampling, trajectory umbrella sampling and optimal reference dynamics}

Since the CDM is a many-body system, as discussed in Sect.~\ref{basic}, accessing the LD statistics of a relevant dynamical order parameter such as the activity for large systems sizes and long times is non-trivial. Sampling of rare trajectories requires a strategy that overcomes the fact that they are exponentially suppressed both in space and in time. This is the central consideration of this paper. 

The first step is to choose a basic scheme to do importance sampling of trajectories, that is, to improve beyond the computational trap of brute force sampling of exponentially rare trajectories with the original dynamics. For this purpose we employ transition path sampling (TPS) \cite{Bolhuis2002}. We consider specifically TPS as formulated for the study of LDs in time-translation invariant dynamics, see e.g.\ \cite{Hedges2009,Oakes2018} (rather than as originally devised to access rare trajectories in transition pathways where the rare events are conditioned to start in one region of phase space and end in another). Alternative numerical methods include ``cloning'' (or ``splitting'') \cite{Giardina2011}, variational approximations \cite{Jacobson2019}, or tensor network approaches \cite{Banuls2019,Helms2019}.  

TPS is Monte Carlo (MC) sampling of trajectories. Like in standard MC there are two relevant basic steps in the TPS algorithm: (i) proposal of trial moves, and (ii) acceptance or rejection. 
For (i) new trajectories are proposed typically via ``shooting'' and/or ``shifting'' moves \cite{Bolhuis2002}, where sections of the current trajectory are regenerated to create the proposed new trajectory. For (ii), a Metropolis rule if often employed, whereby the probability of acceptance is given by $\min{\left( 1, e^{-s \Delta K} \right)}$, where $\Delta K$ is the change in the observable (e.g.\ the activity) between the current and proposed trajectories. This approach is guaranteed to converge to trajectories sampled from \eqref{sens}. 

While TPS (or alternative methods such as cloning) improve the efficiency of sampling with respect to brute force, for many-body systems they may still suffer from slow convergence, especially near LD transitions. To improve convergence one can use umbrella sampling in trajectory space \cite{Ray2018,Ray2018b,Klymko2018,Oakes2018}. If we consider an exponentially tilted average such as the numerator of \eqref{avs} we can write
\begin{align}
\braket{A \ee^{- s K}}
&= \sum_{\omega} \pi_{\rm ref}(\omega) \, A(\omega) \, \ee^{- s K(\omega)} 
\,
\frac{\pi(\omega)}{\pi_{\rm ref}(\omega)} 
\nonumber \\
&= 
\langle A e^{-s K + G} \rangle_{\rm ref} \, ,
\label{us}
\end{align}
where 
\begin{equation}
G(\omega) = \log \frac{\pi(\omega)}{\pi_{\rm ref}(\omega)}  \, .
\end{equation}
This means that we can estimate \eqref{avs} using a different dynamics, the ``reference'', if we exponentially weight according also to the trajectory observable $G$ to account for the change in measure \cite{Ray2018,Ray2018b,Klymko2018,Oakes2018}. A clever choice of reference may reduce
the exponential sampling error. In fact, the optimal choice for the reference dynamics is given via a generalised Doob transformation \cite{Jack2010,Chetrite2015}, which removes the exponential cost of the sampling, basically by cancelling the $-sK$ in the exponent of \eqref{us} with the reweighing term $G$. 

The alternative dynamics has generator 
\beq[W_ref]
\W_{\textnormal{ref}} = \sum_{x,x'\neq x} w^{\textnormal{ref}}_{x \rightarrow x'} \ket{x'} \bra{x} - \sum_{x} R^{\textnormal{ref}}_{x} \ket{x} \bra{x} \, ,
\eeq
with the following restriction: $w^{\textnormal{ref}}_{x \rightarrow x'} \neq 0$ if $w_{x \rightarrow x'} \neq 0$, that is, the reference dynamics connects (with in general different rates) the same configurations as the original dynamics; and, of course, to maintain stochasticity, the escape rates are given by $R^{\textnormal{ref}}_{x} = \sum_{x'} w^{\textnormal{ref}}_{x \rightarrow x'}$.
The reweighing factor for a trajectory $\omega_\tt$ is then 
\begin{align}
e^{G(\omega_\tt)} &= \frac{\pi(\omega_\tt)}{\pi\sub{ref}(\omega_\tt)} 
\nonumber \\
&= \prod_{x,y\neq x}
\left( 
\frac{w_{x \rightarrow x'}}{w^{\textnormal{ref}}_{x \rightarrow x'}}
\right)^{K_{x \rightarrow x'}(\omega_\tt)}
e^{\int_{0}^\tt dt' 
\Delta R_{x_{t'}}
}
\punc,
\end{align}
where $K_{x \rightarrow x'}(\omega)$ indicates the total number of jumps between configurations $x$ and $x'$ in trajectory $\omega$, and $\Delta R_{x_{t'}} = R^{\textnormal{ref}}_{x_{t'}}- R_{x_{t'}}$ with $x_{t'}$ the configuration in the trajectory at time $t'$.
Trajectory averages with the reference dynamics, \eqref{us}, can be computed with TPS by adjusting the acceptance criterion. If the old trajectory was $\omega$ and the proposed one $\omega'$ the probability of acceptance $\Gamma_{\rm acc}(\omega \to \omega')$ reads
\beq[min]
\Gamma_{\rm acc}(\omega \to \omega') = \min\left(1,
\frac{e^{-s K(\omega') + G(\omega')}}
{e^{-s K(\omega) + G(\omega)}}
\right) \, .
\eeq

In the limit of long times the optimal reference dynamics is given by the long-time Doob transformation. This optimal dynamics has generator $\tilde{\W}$ with transitions rates given by 
\beq[WD]
\tilde{w}_{x\to x'} = \frac{\ell_{x'}}{\ell_{x}} \, \ee^{-s} {w}_{x\to x'} ,
\eeq
where the $\ell_x$ are the elements of the leading left eigenvector of $\W_s$, cf.\ \eqref{rls},  and escape rates give by the original ones shifted by the SCGF
\beq[RD]
\tilde{R}_{x} = R_{x} - \theta(s) \, .
\eeq
If the reference dynamics is the one generated by $\tilde{\W}$ the reweighing factor then reads,
\beq[RRD]
e^G = \ee^{\tt \theta(s)} \, \ee^{s K} \, \frac{\ell_{x_0}}{\ell_{x_\tt}} \, .
\eeq
Note that with such reweighing the exponential of $K$ in \eqref{us} is removed, while the exponential of the SCGF cancels the normalisation in \eqref{avs}. In this way the exponential suppression disappears when using the optimal dynamics as the reference to calculate averages such as \eqref{avs}. Note however that the factor $\ell_{x_0}/\ell_{x_\tt}$ which only depends on the time-boundaries of the trajectory remains. For $s \neq 0$ where in general is non-vanishing this will set an upper limit to the efficiency of the sampling with regards to TPS acceptance.

\subsection{Arriving to the optimal dynamics iteratively through feedback}
\label{Secfeed}

The key quantity in defining the optimal dynamics \eqref{WD} is the left eigenvector $\bra{\ell_s}$, cf.\ \eqref{rls}. In practice the problem is that to obtain \eqref{WD} one needs to diagonalise $W_s$ first, something which is often not possible to do in systems of interest. The aim of what follows will be to find a way to estimate the components $\ell_x$ in the most efficient manner.  

The dynamics of the CDM obeys detailed balance which offers a useful simplification. For systems with such dynamics we have in general 
\beq[DB]
P_{\textnormal{eq}}(x) \, w_{x \rightarrow x'} = P_{\textnormal{eq}}(x') \, w_{x' \rightarrow x} \, ,
\eeq
where $P_{\textnormal{eq}}$ is the equilibrium stationary distribution. In operator form this relation can be written as 
\beq[DBop]
\W^{\dagger} = \PP_{\rm eq} \, \W \, \PP_{\rm eq}^{-1} \, ,
\eeq
where $\PP_{\rm eq}$ is a diagonal matrix with entries $P_{\textnormal{eq}}(x)$. For the case of an observable like the activity, the tilted operator also obeys a similar relation
\beq[DBops]
\W_s^{\dagger} = \PP_{\rm eq} \, \W_s \, \PP_{\rm eq}^{-1} \, ,
\eeq
with the same matrix $\PP_{\rm eq}$. 

The optimal dynamics \eqref{WD} is obtained via a generalised (and long time) Doob transform \cite{Jack2010,Chetrite2015},
\beq[Wtilde]
\tW = \LL \, [\W_s - \theta(s)] \, \LL^{-1} \, .
\eeq
The operator $\tW$ obeys $\bra{-} \tW = 0$, as it generates the proper stochastic dynamics \eqref{WD}. The Doob dynamics also obeys detailed balance, 
\beq[DBopt]
\tW^{\dagger} = \tPP \, \tW \, \tPP^{-1} \, ,
\eeq
where $\tPP$ is the diagonal matrix with entries $\tilde{P}(x)$ given by the equilibrium distribution of dynamics \eqref{WD}. 

Combining Eqs.~\eqref{DBops},~\eqref{Wtilde} and \eqref{DBopt} we get 
\beq[L2]
\LL^2 = \PP_{\rm eq}^{-1} \, \tPP \, ,
\eeq
which in components gives the relation between the elements of the left eigenvector and the stationary probabilities of both the original dynamics and the optimal one, 
\beq[feeq]
\ell_x = \sqrt{\frac{\tilde{P}(x)}{P_{\rm eq}(x)}} \, .
\eeq

Equation \eqref{feeq} relates the objects we need to obtain to define the optimal dynamics, $\ell_x$, to two equilibrium distributions. The first one is that of the original dynamics, which we either known exactly - as in the case of the CDM - or we can sample from direct simulation. 
The second one is the equilibrium probability of the Doob dynamics. This one we do not know, but we can {\em estimate} from TPS simulations. This forms the basis of an iterative feedback approach for obtaining the optimal dynamics similar to those used in Refs.~\cite{Nemoto2016,Ferre2018,Das2019}: 
\begin{itemize}
\item We consider a reference dynamics with a structure similar to \eqref{WD}
\beq[Wref]
w^{\rm ref}_{x\to x'} = \frac{\ell^{\rm ref}_{x'}}{\ell^{\rm ref}_{x}} \, \ee^{-s} {w}_{x\to x'} .
\eeq
We could begin for example with $\ell^{\rm ref}_{x} = 1$ for all $x$ as a naive initial guess.

\item We run TPS under this reference and estimate $\tilde{P}$. 

\item We use \eqref{feeq} to obtain a new set of $\ell^{\rm ref}_{x}$, we insert them in \eqref{Wref} to define a new reference dynamics, and repeat the sampling. 
\end{itemize}
Via this procedure the reference dynamics is guaranteed to eventually converge to the optimal Doob dynamics. The practical limitations relate to the size of state space and the ability to estimate the necessary equilibrium densities, which we address in the following sections via machine learning techniques.

\section{Optimal dynamics via Neural networks: supervised learning and network architecture}
\label{sec:NN}

The are several hurdles to overcome in order to implement the iterative feedback procedure of Sect.~\ref{Secfeed}. They all stem from the fact that configuration space grows exponentially with the system size, which in the case of the CDM is given by the number of sites in the lattice. For many-body systems like the CDM the aim is to study the problem for a range of relevant system sizes which immediately leads to a computational bottleneck that is difficult to solve. 

Specifically, for the scheme of Sect.~\ref{Secfeed} we face two related problems. First, in order to define a dynamics \eqref{Wref} one has to be able to specify $\ell^{\rm ref}_{x}$ for all configurations $x$ of the system. As soon as the system exceeds modest sizes there are too many states to tabulate. Secondly, to use \eqref{feeq} to get $\ell^{\rm ref}_{x}$ (that is, an approximation to the true $\ell_{x}$) not only one needs to have an estimate of $\tilde{P}(x)$ for all $x$ - the same problem as for $\ell^{\rm ref}_{x}$ - but furthermore, this estimate has to be constructed from the states sampled in TPS trajectories, which are often much fewer than the total number of states (as is common in sampling). 

These two problems are amenable to solution via standard machine learning methods  \cite{Goodfellow2016,Mehta2019}: the first one we can address via {\em function approximation} whereby we represent the functions $\ell^{\rm ref}_{x}$ and $\tilde{P}(x)$ by means of a neural network (NN) with an overall number of parameters that is much smaller than the dimension of configuration space; the second problem is one of {\em density estimation}, whereby from limited sampled data we approximate the distribution for all possible data. Of course, one can apply functional approximation and density estimation without using NNs  \cite{Goodfellow2016,Mehta2019}, through a choice of approximate description that depends on a small number of parameters. The problem is that if the approximation chosen does not contain the necessary physical information - for example when studying very different dynamical regimes -  then it might have little effect on the sampling process. Using a NN as the approximator mitigates this problem by not hard-constraining physical assumptions.

\begin{figure*}[t]
\begin{center}
\includegraphics[width=\textwidth]{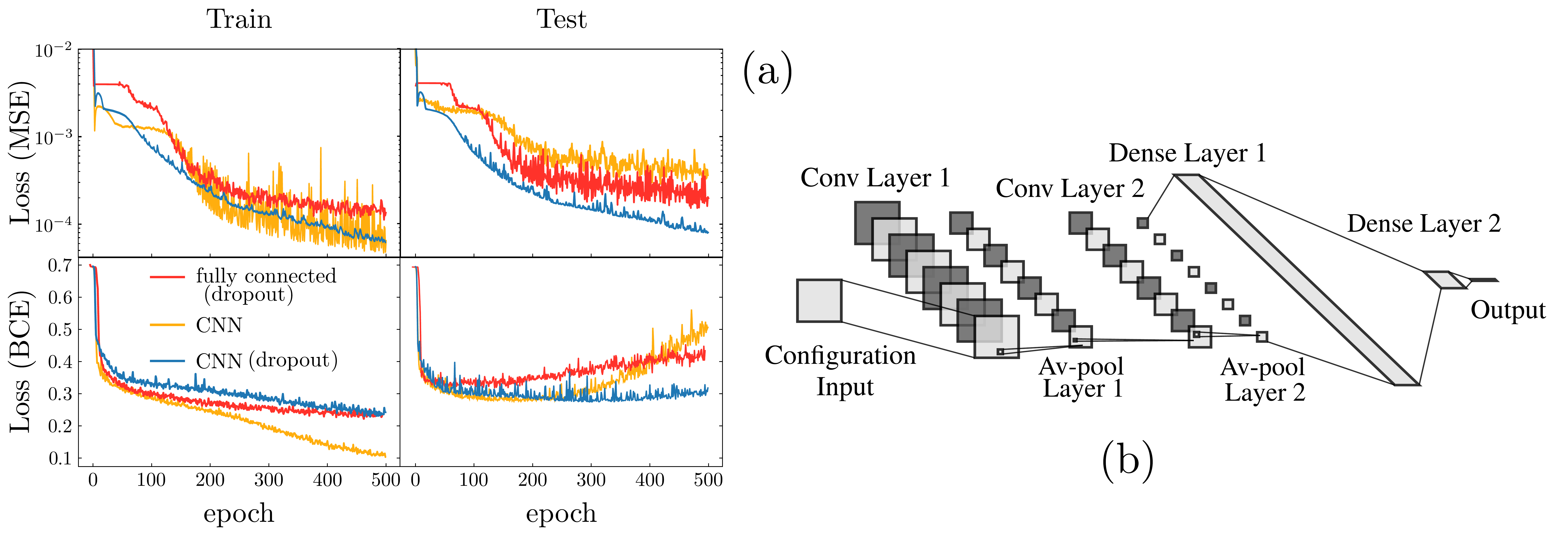}
\end{center}
\caption{(a) Loss of the NNs as a function of epoch. The top panels correspond to the training the NNs to recover the value of $\ell_x$ for each $x$, with the loss defined as the mean squared error (MSE) between the prediction and the exact value (obtained by exact diagonalisation of the titled generator \eqref{Ws}. The bottom panels correspond to training the NNs to make the classification \eqref{mu}, with the loss given by the binary cross entropy (BCE). Shown are three network architectures: fully connected network with 50\% dropout between layers (red), a CNN (orange), and a CNN with 
50\% dropout between layers (blue). The left panels corresponds to the training loss and the right panels to the test loss. Training is done by sampling $10^4$ configurations weighted according to $\ell_x$, while testing is done for $10^3$ sampled configurations. The size of configuration space for the CDM at $N=6 \times 6$ is $\N \sim 10^5$. 
(b) Structure of the NN implemented in the subsequent trajectory sampling. 
For a CDM of size $N = L \times L$ all layers until the first dense layer scale with $L$.
The input is an array of $N = L \times L$ encoding a configuration. The encoding we use consists of allocating a numerical value (1,2,3,4) to the orientation (up, right, down, left) of the half of the dimer at a given site. The two convolutional layers have 8 convolutional nodes each with a kernel size of 2 by 2 and ReLU activation. The two average pooling layers have pooling size of 2 by 2. These are followed by a flattening layer and two dense layers, one with 128 nodes and ReLU activation, the other with 10 nodes and ReLU activation. Between each of the dense layers there is a dropout rate of 50\%.}
\label{fig2}
\end{figure*}

\subsection{Network structure}
\label{subsec:SupeNetwork}

In terms of using a NN, learning how to take a configuration $x$ and map it to a single output $\ell^{\rm ref}_{x}$ can be viewed as a regression problem. If a NN can be trained to provide a value of $\ell^{\rm ref}_{x}$ for each $x$, then the NN can be incorporated in a continuous-time MC algorithm to provide values for the rates of potential moves. 

The optimal reference is the Doob dynamics \eqref{WD}. The first test for our NN approximator is therefore to learn the true form of the Doob transformation vector $\bra{\ell_{s}}$. For CDM this means we are limited to a maximum lattice of size $N=6\times6$ for which we are still able to obtain $\bra{\ell_{s}}$ by exact diagonalisation. The supervised learning problem of training a NN to accurately reproduce the true components $\ell_{x}$ of $\bra{\ell_{s}}$ 
given an input configuration $x$ 
for the exactly solvable $L=6$ case will provide information about the correct network architecture for large scales. 

There are many hyper-parameters associated with finding a suitable network structure. 
Several possible schemes are compared in \reffiga{fig2}{a}. One such is a NN with dense layers and regularisation such as dropout \cite{Goodfellow2016,Mehta2019}. However, if we consider configurations as highly correlated images then it makes sense the use of convolutional layers\cite{Goodfellow2016,Mehta2019}. We can see in \reffiga{fig2}{a} the effect of such layers, in combination with dropout regularisation, has on the training and generalisation of the model. Through this analysis we conclude that the convolutional neural network (CNN) \cite{kashiwa2019, tanaka2017, efthymiou2019} shown in \reffiga{fig2}{b} is an appropriate architecture for a functional approximator of $\ell_x$. 

An important step is an adaptation of the loss function which will become very useful when we scale the problem up in the next section. 
The key expression for what follows is \eqref{feeq}. What this equation shows is that in order to obtain the vector $\ell_x$ that defines the reference dynamics we are not interested in the individual probability distributions, $P_{\rm eq}$ and $\tilde{P}$, but on the ratio between them for each configuration $x$. This fact can be exploited using a method of learning that rephrases the issue of density estimation as a semi-supervised learning problem \cite{Friedman2001}. 

Specifically, consider a situation where we are generating sample configurations $x$ from two distributions, the equilibrium one $P_{\rm eq}$ corresponding to the original dynamics \eqref{W}, and a second stationary distribution $\hat{P}$, for example an estimate of the true Doob equilibrium $\tilde{P}$. Consider the problem of training a NN to discriminate between configurations generated from $P_{\rm eq}$ or $\hat{P}$. That is, the output of a NN such as that of Fig.~\ref{fig2}(b) could be a function $\mu(x) \in [0,1]$ trained to reproduce 
\begin{equation}
\label{mu}
\mu(x) = 
\frac{\hat{P}(x)}{\hat{P}(x) + P_{\rm eq}(x)} = 
\frac{\frac{\hat{P}(x)}{P_{\rm eq}(x)}}{1 + \frac{\hat{P}(x)}{P_{\rm eq}(x)}} \, .
\end{equation}
The above expression corresponds to the average of a binary label $Y$ which is equal to $Y=1$ if $x$ is generated exclusively by $\hat{P}$ and $Y=0$ if $x$ is generated exclusively by $P_{\rm ref}$. This is a standard classification problem. We can define our NN Fig.~\ref{fig2}(b) with a sigmoid activation function before the output, corresponding to $\mu(x)$, and train it with data obtained in an i.i.d.\ from both $\hat{P}$ and $P_{\rm eq}$ and labelled accordingly. Furthermore, from the output $\mu(x)$, we can obtain an estimate of the left eigenvector $\ell_x$ from the output of the NN by rearranging \eqref{mu},
\beq[lm]
\ell_x = \frac{\mu(x)}{1-\mu(x)} \, .
\eeq

\begin{figure*}[t]
\begin{center}
\includegraphics[width=\textwidth]{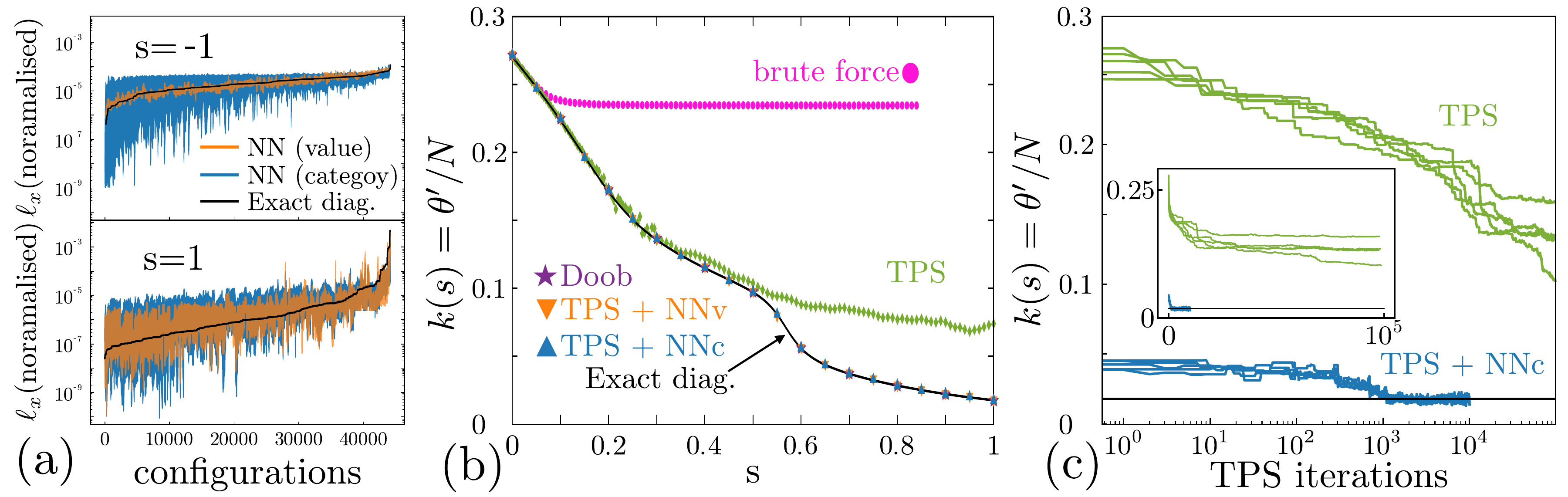}
\end{center}
\caption{(a) Value of $\ell_x$, normalised by $\sum_x \ell_x$, for all configurations $x$, for the CDM at $N=6\times 6$. The top panel is for $s=-1$ while the bottom panel for $s=1$. The plots show the normalised values of $\ell_x$ from exact diagonalisation of \eqref{Ws} (orange), and for the NNs of Fig.~\ref{fig2}(b) with output the vector $\ell_x$ (green), and with the output the classification \eqref{mu} (blue). The configurations $x$ are ordered based on ascending values of the exact $\ell_x$. 
(b) Activity per unit time and size in the tilted ensemble, $k(s) = \langle K \, e^{-sK} \rangle / (\langle e^{-sK} \rangle \, N \, t) = -\theta'(s)/N$, vs $s$ for the case of $=6 \times 6$. The pink circles show $k(s)$ estimated by brute force (from $10^5$ sample trajectories for each $s$ followed by exponential reweighing). 
The green diamonds correspond to standard TPS (starting at $s=0$ and moving in steps of $\delta s = 10^{-2}$ after $10^5$ TPS iterations, and similarly for the other TPS data).
The down blue triangles correspond to TPS with a reference dynamics where the NN trained for the value of $\ell_x$ (NNv) is used as function approximator. The up orange triangles are the same but with the NN trained with the category approach (NNc). They both coincide with the use of the exact diagonalisation result (black curve), and with TPS with the exact Doob dynamics \eqref{WD} (purple stars) as reference (TPS needed in this case to fix time-boundary effects). 
(c) Sample TPS evolution collected at $s=1$ from the standard TPS method (blue) and from TPS with reference dynamics given by the trained NNc (orange). The inset is the same data on a linear scale.}
\label{fig3}
\end{figure*}

\subsection{Proof of principle: LD sampling for $N=6 \times 6$}
\label{subsec:L=6Results}

As a proof of principle on the use of a NN for the function approximator of the reference dynamics we consider the CDM for size $N=6 \times 6$ and periodic boundaries, which is solvable by exact diagonalisation. In this case we can calculate exactly the vector $\ell_x$ for all configurations $x$. Figure \ref{fig3}(a) shows the ability of a NN like that of Fig.~\ref{fig2}(b) to learn the exact $\ell_x$. We show for comparison the NN trained to reproduce the classification \eqref{mu} (note that for the CDM the equilibrium probability $P_{\rm eq}$ is uniform over all configurations), with a similar network trained to reproduce the actual value of $\ell_x$. While the classification/category NN (NNc) gives larger fluctuations under training than the value NN (NNv), as we will see below, it behaves better under generalisation. 

The aim is to use this network within a TPS simulation as an aid to define the reference dynamics. Figure \ref{fig3}(b) compares the results from various sampling schemes with the exact diagonalisation of the tilted generator \eqref{Ws} for the $N=6 \times 6$. We plot the average activity per unit time 
\begin{equation}
k(s) = -\theta'(s) \,
\end{equation}
against $s$, cf.~\cite{Oakes2018}. The sampling data is shown for the same number of generated trajectories ($10^5$) for each $s$, in order to compare their efficiency. 
The pink circles show brute force sampling via post-selection: as expected, the exponential cost of sampling away from $s=0$ makes the estimation of the true $k(s)$ very poor. 
The green diamonds are results obtained via standard TPS, where trajectories are generated using the original dynamics: while this is an improvement over brute force sampling, for $s$ large enough the simulation fails to reproduce the exact values (black curve). 
The down blue triangles and up orange triangles correspond to TPS with a reference dynamics provides by NNs trained with the value (NNv) and category (NNc) of $\ell_x$. 
They coincide with TPS using the exact Doob dynamics \eqref{WD} as reference (purple stars), and with the exact diagonalisation result. Figure  \ref{fig3}(b) demonstrates that TPS supplemented by appropriately trained NNs [with the architecture of Fig.~\ref{fig2}(b)] can efficiently recover the dynamical LD statistics of the CMD, at least for $N = 6 \times 6$. The next section addresses how to scale up this approach.

As an alternative measure of efficiency we can compare the number of TPS iterations required to equilibrate (in the TPS sense) to the correct value of $k(s)$. Figure \ref{fig3} shows such comparison between standard TPS and TPS supplemented by the NNc for $s=1$. We see from the figure that for these conditions the NNc enhanced TPS converges to the exact value orders of magnitude faster than standard TPS.

\begin{figure*}[t]
\begin{center}
\includegraphics[width=\textwidth]{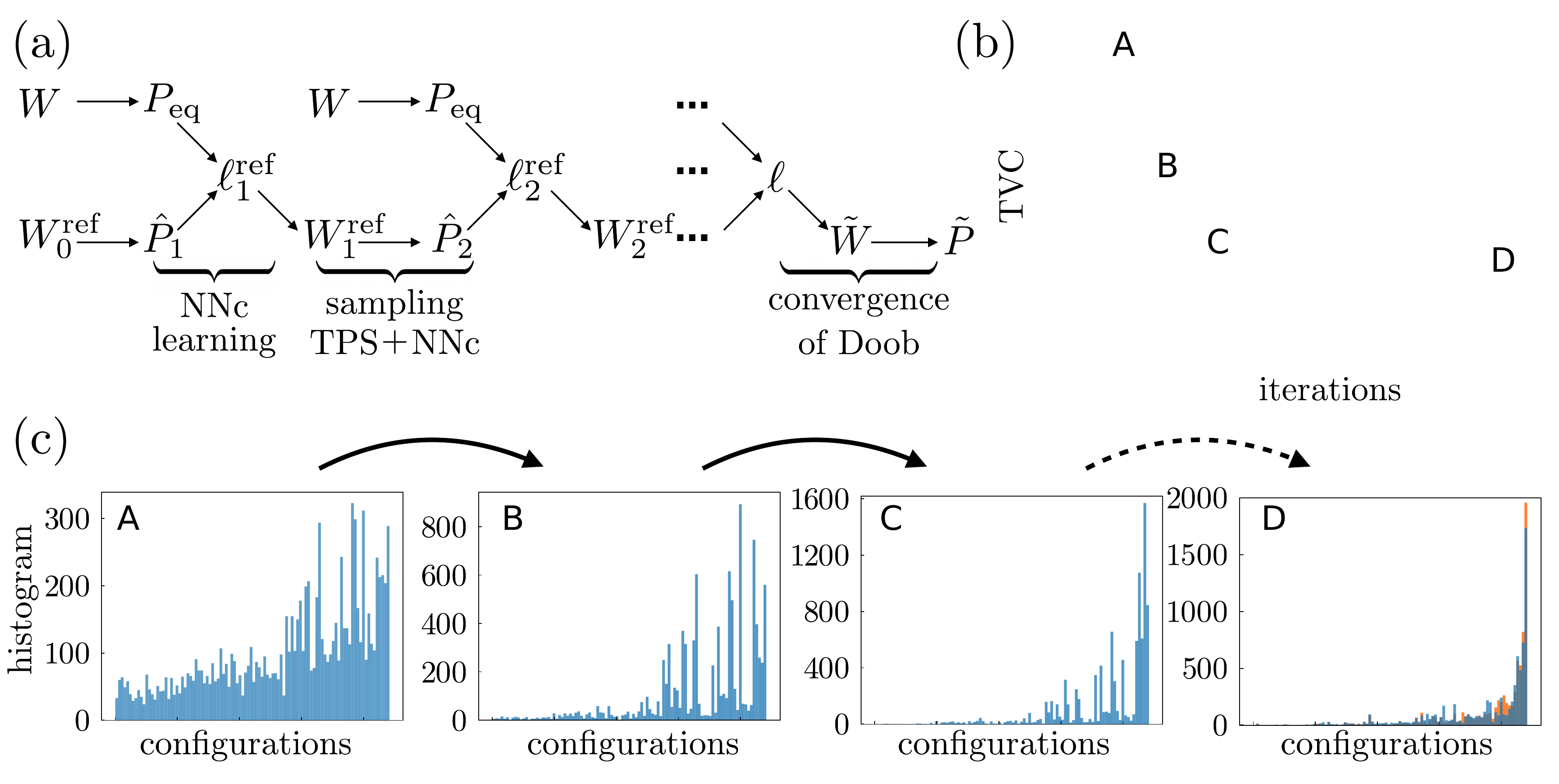}
\end{center}
\caption{(a) 
Sketch of the iterative procedure to arrive at accurate sampling of the tilted ensemble. Each iteration of the procedure has two stages. The first one corresponds to sampling the equilibrium probability with the original dynamics, \eqref{W}, and the probability of the tilted ensemble via TPS with a reference dynamics, \eqref{Wref}, which at iteration $k$ is  
$\mathbb{W}^{\rm ref}_{n-1}$ (for the initial iteration we choose $\ell^{\rm ref}_x$ uniform). At the end of this stage we obtain samples of an approximation $\hat{P}_k$ to the true $\tilde{P}$. The second stage of the iteration corresponds to training the NN to classify the samples, \eqref{mu}. This via \eqref{lm} allows to generate via the NN the vector $\ell^{{\rm ref},n}_x$, which is used to define the reference dynamics for the next iteration 
$\mathbb{W}^{\rm ref}_{n}$. This feedback scheme should eventually converge
to the exact Doob vector $\ell$ and corresponding optimal dynamics \eqref{WD}. 
(b)
Total variational distance between the (normalised) exact $\ell$ (for $N = 6\times 6$ and $s=1$) and the output of the NNc learned at each stage of the iterative procedure (green symbols). The orange dashed line indicates the TVD between the exact $\ell$ and the NNc trained with the exact values, cf.\ Figs.~\ref{fig2},\ref{fig3}, while the blue dotted line the same for NNv. 
(c) Evolution of the probability distribution in the iterations labelled in the previous panel. In iteration D we compare to the exact value (orange). 
}
\label{fig4}
\end{figure*}

\section{
Optimal dynamics via Neural networks: Semi-supervised learning and scalability}
\label{sec:Unsuper}

\subsection{Iterative feedback procedure}

For system sizes beyond $N=6 \times 6$ it is not possible to numerically diagonalise the tilted generator of the CMD. In order to scale up the procedure of the last section we 
can adapt the optimal control procedure introduced in \cite{Nemoto2016} (see also \cite{Ferre2018,Das2019}) to our setting, as discussed above in Sect.~\ref{Secfeed}. Figure \ref{fig4}(a) sketches our feedback scheme. The procedure consists of two stages per iteration. The first one samples both the equilibrium configurations from $P_{\rm eq}$ with the original dynamics \eqref{W} (something which is redundant in the CMD as the equilibrium distribution is uniform), and configurations in the ensemble tilted by $s$, using TPS and a reference dynamics. This latter step is imperfect, and only produces samples according to an approximation $\hat{P}$ to the true target distribution $\tilde{P}$. The second stage of each iteration is to train the NN to classify between configuration generated by $P_{\rm eq}$ and $\tilde{P}$. The NN we use has the structure of that of Fig.~\ref{fig2}(b). The trained network is then used to generate the vector $\ell^{\rm ref}$ that determines the reference dynamics, cf.~\eqref{Wref}, for the next iteration of the procedure. For enough iterations we eventually converge to the optimal dynamics \eqref{WD}. 

We can test the effectiveness of this iterative procedure by looking at the probability distribution produced by each step in the iteration. Figure 
\ref{fig4}(b) shows the 
distance between the left eigenvector $\ell^{{\rm ref},n}_x$ at each iteration $n$ the feedback procedure and the exact one $\ell_x$. If we normalise these vectors we obtain probability distributions over the configurations and we can compare them in terms of total variational distance (TVD)
\begin{equation}
{\rm TVD}(\ell^{{\rm ref}},\ell) 
= 
\sum_{x} \left|\frac{\ell^{{\rm ref}}_x}{\sum_{y}\ell^{{\rm ref}}_y} - \frac{\ell_x}{\sum_{y}\ell_y} \right|.
\end{equation}
We see from Figs.~\ref{fig4}(b,c) for the solvable $N=6\times6$ case that over a small number of iterations the NNs learn the values as well as they would have done so using the exact diagonalisation data as input.

\begin{figure*}[t]
\begin{center}
\includegraphics[width=\textwidth]{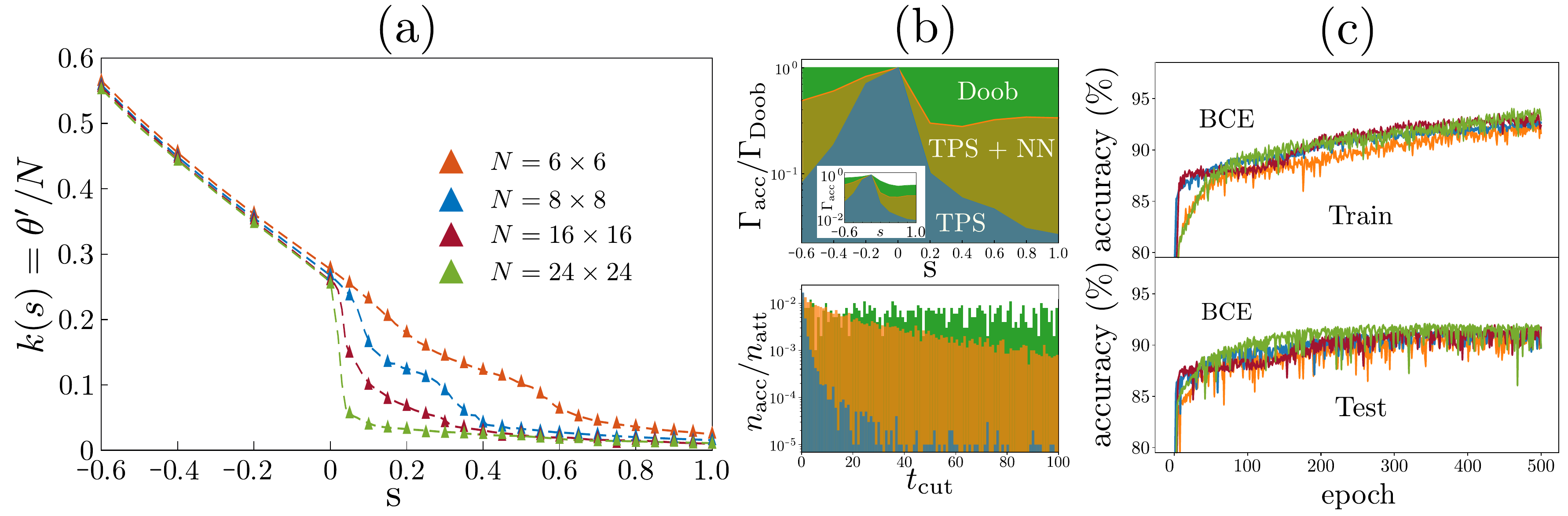}
\end{center}
\caption{
(a)
Average activity as a function of $s$. Symbols are for systems sizes $N = 6\times6$ (orange),  $N = 8\times8$ (blue), $N = 16\times16$ (red) and $N = 24\times24$ (green), obtained from TPS with the NNc reference via the feedback procedure, cf.\ Fig.~\ref{fig4}. The dashed lines correspond to the results of Ref.~\cite{Oakes2018}. 
(b) In both panels the system size is $N = 6\times6$.
Top panel: TPS acceptance rate as a function of $s$. When the original dynamics is used (blue) acceptance decays fast away from $s=0$. Acceptance is much larger with the reference obtained from NNc (orange). In the main plot we show acceptance scaled by that of the optimal Doob dynamics (green; the inset gives the unscaled rate, showing that even with the optimal dynamics acceptance is not strictly unity away from $s=0$ due to time-boundary effects). 
Bottom panel: TPS acceptance rate at $s=1$ as a function of the time of the cut. A good choice of reference dynamics reduces the exponential suppression with time of the acceptance. 
(c)
The accuracy of NN trained on the binary-cross entropy try to distinguish two distributions, cf.~\eqref{mu}. The top panel shows the binary cross entropy for training and the bottom one for test data, for the same system sizes of (a).
}
\label{fig5}
\end{figure*}

\subsection{Sampling Results}
\label{subsec:L>6Results}

We are now in a position to use the semi-supervised learning method to sample systems that are not accessible to exact diagonalisation. Figure \ref{fig5}(a) shows the average activity per unit time and size, $k(s)/N$, as a function of $s$ for different system sizes in the CMD. We see from the figure that the results coincide with those of Ref.~\cite{Oakes2018}, currently the state of the art for this model. The key observation of \cite{Oakes2018} was the existence of a transition at the LD level at $s=0$ in the limit of large system size, seen as the change in $k(s)$ that is tending to a discontinuity. Accurate sampling of trajectories responsible for the transition is the main computational difficulty in this system. The results of  Figure \ref{fig5}(a) show that our NN assisted feedback procedure is capable of sampling such rare events efficiently. 

One caviat of this process, arose from the choice of initial dynamics for the largest system size we explored, $N = 24\times24$. In the smaller system sizes we simply set $\ell^{\rm ref}_x$ to be uniform as an initial dynamics. However for $N = 24\times24$, the distribution produced by setting $\ell^{\rm ref}_x$ to be uniform was not sufficently distinct from the equilibrium distribution for the iterative procedure to converge in a resonable time. To solve this problem we explored the use of transfer learning by Training a NN on a tiled version of a smaller system size, $N = 6\times6$; such that the input to the NN is sufficently sized for $N = 24\times24$ configurations. The inital dynamics can be implimented using the trained NN instead of the uniform $\ell^{\rm ref}_x$. The initial distribution produced from the trained NN was then capable of converging in a reasonable time.

The effectiveness of the iterative NN method can be seen in the ability to find a reference dynamics under which to run TPS where the acceptance of proposed trajectory moves is larger than in standard TPS. In Fig.~\ref{fig5}(b) (top panel) we show the TPS acceptance as a function of $s$. For TPS where trajectories are generated with the original dynamics, this acceptance decreases fast with $s$. Furthermore, as shown in Fig.~\ref{fig5}(b) (bottom panel) for the $s=1$ case, acceptance is exponentially suppressed with the time point at which the current trajectory is cut to generate the proposed one. In contrast, TPS with the NN reference dynamics has a much larger overall acceptance probability away from $s=0$ and the exponential reduction with cut time is much suppressed. In these sense it is getting closer to the optimal Doob dynamics, for which acceptance probability is in principle independent of cut time [but not quite equal to unity away from $s=0$ due to the need to correct the temporal boundaries in trajectories, see discussion following Eq.~\eqref{RRD}].

A final check on how well the semi-supervised NN method is performing, instead of looking at the loss function of the NN under training/testing, we can consider the accuracy
as shown in Fig.~\ref{fig5}(c). This is defined as the percentage of correctly categorised configurations, and is a metric that can be applied to any system size. We use the exact data from $N=6\times6$ as a benchmark.

\section{Conclusions}
\label{Secconc}

We have addressed the problem of numerically sampling rare dynamical trajectories for the estimation of the large deviation statistics of dynamical observables. We have considered this problem specifically for the case of the classical dimer model as a characteristic many-body system where the sampling complexity arises from its correlated dynamics and associated trajectory phase transition. We have shown that it is possible to implement a scheme where the optimal dynamics with which to sample exponentially tilted trajectory ensembles is obtained from an iterative feedback mechanism with minimal input regarding the nature of the problem. Specifically, by using a neural network as a functional approximator for the eigenvector over the many-body configurations that defines the reference dynamics, we are able to implement our scheme in a manner that is scalable with system size. For the CMD we showed that this procedure efficiently recovers the LD phase transition of the model. Our approach should be directly applicable to other many-body systems of interest. Our results are yet another example of the broad potential of deep learning methods for study of non-equilibrium systems more generally.

\acknowledgements

We are grateful to Ed Gillman for important discussions. This work was supported by an EPSRC Doctoral Prize (THEO) from the
University of Nottingham, a Royal Society University Research Fellowship (AM) and an EPSRC Grant No.\ EP/R04421X/1 (JPG). We are greatful for access to the University of Nottingham's Augusta HPC service. We also acknowledge the use of Athena at HPC Midlands Plus.

\bibliography{MLtps}

\bibliographystyle{apsrev4-1} 

\end{document}